\documentclass[12pt]{iopart}
\usepackage{iopams}  

\usepackage{setstack}
\usepackage{cite}
\usepackage{graphicx}
\usepackage{hyperref}
\usepackage[usenames,dvipsnames]{xcolor}
\hypersetup{colorlinks=true, linkcolor=BrickRed, urlcolor=blue!50!black, citecolor=blue!50!black}

\usepackage[normalem]{ulem}

\begin{document}

\title[Dynamic properties of quasi-confined colloidal liquids]{Dynamic properties of quasi-confined colloidal hard-sphere liquids near the glass transition}

\author{Lukas Schrack, Charlotte F. Petersen, Gerhard Jung, Michele Caraglio, and Thomas Franosch}

\address{Institut f\"ur Theoretische Physik, Universit\"at Innsbruck, Technikerstr. 21A, 6020 Innsbruck, Austria}
\ead{lukas.schrack@uibk.ac.at}
\vspace{10pt}
\begin{indented}
\item[]\today
\end{indented}

\begin{abstract}
The complex behavior of confined fluids arising due to a competition between layering and local packing can be disentangled by considering \emph{quasi-confined liquids}, where periodic boundary conditions along the confining direction restore translational invariance. This system provides a means to investigate the interplay of the relevant length scales of the confinement and the local order. We provide a mode-coupling theory of the glass transition (MCT) for quasi-confined liquids and elaborate an efficient method for the numerical implementation. The  nonergodicity parameters in MCT are compared to computer-simulation results for a hard-sphere fluid. We evaluate the nonequilibrium-state diagram and investigate the collective intermediate scattering function. For both methods, nonmonotonic behavior depending on the confinement length is observed. 
\end{abstract}

%
\vspace{2pc}

\submitto{\JSTAT}
%
%
%

\section{Introduction}
Confining a liquid introduces competition between near-range local ordering by the cage effect and constraints due to the boundaries. These confined liquids bridge the gap between 3D bulk liquids and quasi-2D systems. They are ubiquitous in nature, for instance in biosystems or geological processes and have important industrial applications, ranging from chemical synthesis and bioanalysis to optics and information technology~\cite{Urbakh:Nature:2004, Whitesides:Nature:2006}. 
Confinement influences most of the structural and dynamical properties of a fluid~\cite{Loewen:JPCM:2001, Alba:JoP:2006, Varnik:JP:2016}, in particular diffusion~\cite{Mittal:PRL:2008}, the freezing transition~\cite{Schmidt:PRL:1996, Schmidt:PRE:1997} and the glass transition~\cite{ Krakoviack:PRL:2005, Krakoviack:JoP:2005, Krakoviack:PRE:2007, Krakoviack:PRE:2011, Kim:EPL:2009, Lang:PRL:2010, Szamel:EPL:2013}, where structural arrest exceeds macroscopic time scales.

One of the simplest realizations of confinement is a slit geometry consisting of two parallel flat walls. This system has been investigated extensively in experiments~\cite{Nugent:PRL:2007, Edmond:PRE:2012, Sarangapani:PRE:2011, Sarangapani:SoftMatter:2012, Eral:PRE:2009, Eral:Langmuir:2011,  Nygard:PRL:2012, Nygard:JCP:2013, Nygard:PRX:2016, Nygard:PRL:2016, Nygard:PCCP:2017, Nygard:PRE:2017} and simulations~\cite{Fehr:PRE:1995, Scheidler:EPL:2000, Scheidler:EPL:2002, Scheidler:JoPCB:2004, Torres:PRL:2000, Varnik:PRE:2002, Varnik:JoCP:2002, Baschnagel:JoP:2005, Mittal:PRL:2006, Mittal:JoPCB:2007, Mittal:JoCP_127:2007, Mittal:PRL:2008, Krishnan:JoCP:2003, Krishnan:PRE:2012, Varadarajan:JCP:2018, Goel:PRL:2008, Goel:JStatMech:2009, Krekelberg:JoCP:2011, Krekelberg:Langmuir:2013, Krekelberg:Langmuir:2017, Deb:JoCP:2011, Ingebrigsten:PRL:2013, Ingebrigtsen:SoftMatter:2014, Mandal:NatComm:2014, Geigenfeind:JoCP:2015, Saw:JoCP:2016, Bollinger:SoftMatter:2016, Ghosh:PRE_97:2018, Ghosh:PRE_98:2018, Ghosh:SciRep:2019}.
The slit introduces an additional length scale as a control parameter, which in the limit of strong confinement rivals the typical interaction range. The competition between local packing induced by the cages of neighboring particles and layering induced by the walls results in a nonmonotonic behavior of the diffusivity~\cite{Mittal:PRL:2006, Mittal:PRL:2008, Goel:PRL:2008, Bollinger:JCP:2015}  and the glass transition on the wall distance~\cite{Lang:PRL:2010, Mandal:NatComm:2014}. In recent theoretical studies the decoupling between transverse and lateral degrees of freedom in the limit of strong confinement was used to gain a deeper insight into the structure~\cite{Franosch:PRL:2012, Lang:JCP:2014} and dynamics~\cite{Schilling:PRE:2016, Mandal:PRL:2017, Mandal:EPJST:2017} of strongly confined liquids.

Within these liquids it has been found empirically that transport properties correlate with purely thermodynamic properties such as the excess entropy~\cite{Mittal:PRL:2006, Mittal:PRL:2008, Goel:PRL:2008, Ingebrigsten:PRL:2013, Bollinger:JCP:2015, Ingebrigtsen:PNAS:2018}. In particular, the quasi-universality of simple liquids has been investigated extensively~\cite{Ingebrigtsen:PRX:2012, Dyre:JoP:2016}. A recent review on excess-entropy scaling is provided in Ref.~\cite{Dyre:JoCP:2018}.

Quite similarly, the mode-coupling theory of the glass transition (MCT)~\cite{Goetze:Complex_Dynamics} predicts the dynamical behavior using only structural information as input. 
It rationalizes many nontrivial facets of the glass transition~\cite{Goetze:Complex_Dynamics, Sperl:PRE:2000, Voigtmann:PRL:2009, Voigtmann:EPL:2011, Gnan:PRL:2014, Janssen:FiP:2018}, in particular, the existence of a  structural arrest where the dynamics change from ergodic to nonergodic behavior in spite of the static quantities varying only smoothly  at this point. In the vicinity of the glass transition MCT predicts a two-step structural relaxation~\cite{Goetze:Complex_Dynamics,Franosch:PRE_55_6:1997}, where the first scaling law describes the dynamics close to the plateau value (indicated by the nonergodicity parameter), and the second one explains the decay to zero by stretched relaxation functions.

Using symmetry-adapted modes MCT has been successfully extended to confined liquids within a slit geometry for Newtonian dynamics~\cite{Lang:PRL:2010, Lang:PRE:2012, Lang:PRE_89:2014, Lang:PRE_90:2014, Lang:JStatMech:2013,Jung:JStatMech:2020,Jung:2020}, and only recently also for Brownian microscopic dynamics~\cite{Schrack:PhilMag:2020}.

The numerical evaluation of MCT for confined fluids is rather involved since the fluid becomes inhomogeneous, and as such many couplings between the symmetry-adapted modes have to be considered. In addition, the confining walls lead to an interplay of layering  and  local packing, and it remains difficult to identify which ingredient dominates the transport properties. Here, we circumvent both of these problems by investigating quasi-confined liquids,  where periodic boundary conditions in the confining direction are employed. This model is translationally invariant in the confining direction as well as in the unconfined directions, implying that the density profile is uniform. Yet, the particles are still significantly affected by the confinement when the confinement length becomes comparable to the particle diameter. Therefore, confinement effects can be studied without the complexity due to walls or density modulations. This enables us to disentangle confinement and layering effects, that appear to be intimately tied, and to determine the dominant contribution. These fluids are a special case of the more general class of layered fluids discussed above. Here, the dynamics are restricted to the surface of a four-dimensional cylinder. A similar approach has been applied for studying the nonequilibrium dynamics in narrow channels using confined lattices~\cite{Benichou:PRL:2013, Benichou:PRE:2016}.

Recently, we have elucidated  the static properties of quasi-confined liquids using integral theory and event-driven simulations~\cite{Petersen:JStatMech:2019}. Here,  we extend these investigations to the dynamical behavior. We start with a theoretical description of the relevant MCT equations adapted to quasi-confined liquids in~\sref{sec:theoretical_description} and demonstrate an efficient strategy for the solution by introducing an effective memory kernel.
The details about the simulations are described in~\sref{sec:simulation}. In~\sref{sec:result} we elaborate the similarities and differences of the MCT calculations and computer simulations. Finally, in~\sref{sec:conclusion} we provide a critical assessment of our results, in particular, in comparison to the slit geometry. Technical details regarding the numerical solution of the equations of motion are described in~\ref{sec:appendix_numerics}, simulation results for longer simulation times in ~\ref{sec:appendix_ISF}.

\section{Theoretical description}\label{sec:theoretical_description}
\subsection{Equations of motion}
We examine a 3D colloidal suspension of identical hard spheres with diameter $\sigma$ undergoing overdamped Brownian motion, ignoring hydrodynamic interactions. Applying  periodic boundary conditions in the confining direction implies that this dimension is compactified, and the dynamics of the suspension can thus be considered to occur on the surface of a four-dimensional cylinder. In the thermodynamic limit, $N\to\infty$, $A\to\infty$ the area density $n_0=N/A$ and the volume density $n=n_0/L$ remain constant, where $A$ denotes the  area in the dimensions perpendicular to the confinement. The packing fraction of the system is then given by $\varphi=n\pi\sigma^3/6$.

We denote the in-plane or lateral coordinates by $\vec{r}=(x,y)$ and the transversal periodic dimension is identified with $-L/2\le z \le L/2$. In the thermodynamic limit both lateral coordinates $x$ and $y$ extend to infinity. We introduce the short-hand notation $\vec{x}=(\vec{r},z)$ for positions on the surface of the four-dimensional cylinder.

These quasi-confined liquids constitute  a special case of layered fluids characterized by translational symmetry along the $x$-$y$-direction and rotational symmetry around the $z$-axis. The quasi-confinement additionally implies translational symmetry in the  $z$-direction. Static properties of quasi-confined liquids have been studied recently~\cite{Petersen:JStatMech:2019}. In this section we translate the mode-coupling equations from the more general case of a fluid within a slit geometry~\cite{Lang:PRL:2010,Lang:PRE:2012}, adapted to  Brownian dynamics~\cite{Schrack:PhilMag:2020}, by employing the additional symmetries due to the restored translational symmetry. 

The key quantity in our discussion is the collective intermediate scattering function (ISF)
\begin{equation}\label{eq:ISF}
 S_\mu(q,t)=\frac{1}{N}\left\langle\rho_\mu(\vec{q},t)^*\rho_\mu(\vec{q})\right\rangle,
\end{equation}
where the mode index $\mu\in\mathbb{Z}$ corresponds to a discrete wavenumber $Q_\mu=2\pi\mu/L$ and $\vec{q}=(q_x,q_y)$ is the continuous wavevector in the lateral direction. The initial value $S_\mu(q,t=0)=: S_\mu(q)$ is given by the static structure factor of the quasi-confined liquid~\cite{Petersen:JStatMech:2019}. The ISF can be identified  with the diagonal element of the generalized ISF within the slit geometry, $S_\mu(q,t)\equiv S_{\mu\mu}(q,t)$~\cite{Petersen:JStatMech:2019}. Thereby, symmetry-adapted microscopic fluctuating density modes
\begin{equation}\label{eq:denmodes}
 \rho_{\mu}(\vec{q})=\sum_{n=1}^N\exp{\left[\rmi Q_\mu z_n\right]}e^{\rmi \vec{q}\cdot\vec{r}_n},
\end{equation}
have been introduced. 

The collective ISF for quasi-confined liquids naturally depends only on a single mode index $\mu$ due to translational invariance along the $z$-direction, in contrast to the two mode indices appearing for the slit geometry.  Furthermore, due to mirror reflection symmetry the correlators do not depend on the sign of the mode index, $S_\mu(q,t)=S_{-\mu}(q,t)$.
The equations for quasi-confined liquids are reminiscent to the diagonal approximation~\cite{Lang:PRL:2010,Lang:PRE:2012,Jung:2020} for the slit geometry and there is no coupling of the ISF for different mode indices. Nevertheless, our system is different from the diagonal approximation of the slit geometry. First, for quasi-confined liquids it becomes an exact symmetry rather than a technical approximation. Second, the static input differs  due to the inhomogeneous density profile for layered fluids compared to the constant density for quasi-confined liquids~\cite{Petersen:JStatMech:2019}. Third, quasi-confined  liquids include couplings between different relaxation channels which are ignored within the diagonal approximation for the slit geometry.

Using the Zwanzig projection operator formalism~\cite{Goetze:Complex_Dynamics, Forster:Hydrodynamic_Fluctuations} the exact equations of motion (e.o.m.) for the collective correlator $S_\mu(q,t)$ read
\begin{eqnarray}\label{eq:eom_correlator}
 \fl\dot{S}_{\mu}(q,t) + D_{\mu}(q)S_{\mu}(q)^{-1}S_{\mu}(q,t)+ \int_0^t\delta K_{\mu}(q,t-t^\prime) S_\mu(q)^{-1} S_{\mu}(q,t^\prime)\mathrm{d}t^\prime = 0.
\end{eqnarray}
The initial decay of the correlator is then given by
\begin{equation}
 D_\mu(q)=(q^2+Q_\mu^2) D_0,
\end{equation}
with the  bare diffusion coefficient $D_0$.
The memory kernel $\delta K_\mu(q,t)$ plays the role of a generalized friction coefficient reflecting the influence of all other modes on $S_\mu(q,t)$. A crucial feature of the theory is the fact that the memory kernel $\delta K_\mu(q,t)$ naturally splits into relaxation channels parallel and perpendicular to the confinement direction
\begin{eqnarray}
 \delta K_\mu(q,t)=\sum_{\alpha\beta=\parallel,\perp}b^\alpha(q,Q_\mu)\delta\mathcal{K}_\mu^{\alpha\beta}(q,t)b^\beta(q,Q_\mu),
\end{eqnarray}
with channel indices $\alpha,\beta \in \{ \parallel, \perp \}$ and selector $b^\alpha(x,z)=x\delta_{\alpha\parallel}+z\delta_{\alpha\perp}$. An analogous splitting can be achieved for the diffusion coefficient with channel diffusion matrix $\mathcal{D}_\mu^{\alpha\beta}(q)=\delta_{\alpha\beta} D_0$.

\subsection{Irreducible memory kernel}
The matrix-valued memory kernel $\delta\mathcal{K}_\mu^{\alpha\beta}(q,t)$ can be expressed in terms of an irreducible memory kernel which is more suitable for applying MCT approximations~\cite{Cichocki:PhysicaA:1987, Kawasaki:PhysicaA:1995}. The concept of irreducible memory functions is also present in modified MCT approaches, e.g.\ within the self-consistent generalized Langevin equation theory~\cite{Yeomans-Reyna:PRE_64:2001, Yeomans-Reyna:PRE_76:2007} or within a field-theoretic self-consistent perturbation approach~\cite{Kim:JStatMech:2008,Kim:PRE:2014}. Then the matrix-valued e.o.m.\ for the memory kernel are given by
\begin{eqnarray}\label{eq:eom_irreducible}
 \fl\delta\boldsymbol{\mathcal{K}}_\mu(q,t)=-\boldsymbol{\mathcal{D}}_\mu(q)\boldsymbol{\mathcal{M}}_\mu(q,t)\boldsymbol{\mathcal{D}}_\mu(q)-\int_0^t \boldsymbol{\mathcal{D}}_\mu(q)\boldsymbol{\mathcal{M}}_\mu(q,t-t^\prime)\delta\boldsymbol{\mathcal{K}}_\mu(q,t^\prime) \mathrm{d}t^\prime,
\end{eqnarray}
with the irreducible memory kernel $\boldsymbol{\mathcal{M}}_\mu(q,t)$. Here we employ a matrix notation in the channel index, i.e. $[\delta\boldsymbol{\mathcal{K}}_\mu(q,t)]^{\alpha\beta} = \delta\mathcal{K}_\mu^{\alpha\beta}(q,t)$, the products are to be understood as matrix multiplications. The e.o.m., equations~\eref{eq:eom_correlator} and~\eref{eq:eom_irreducible}, are then closed using 
suitable MCT approximations by writing the irreducible memory kernel as a bilinear functional of the ISF,
\begin{equation}\label{eq:MCT_functional}
 \mathcal{M}_\mu^{\alpha\beta}(q,t) =\mathcal{F}_\mu^{\alpha\beta}\left[S(t),S(t);q\right],
\end{equation}
where $S(t)$ abbreviates the collection of the ISF for all possible mode indices and wavenumbers. It has been shown recently that by a proper choice of the irreducible (adjoint) Smoluchowski operator the explicit expression for the force kernel is identical for  Newtonian and Brownian microscopic dynamics~\cite{Schrack:PhilMag:2020}.

Due to the splitting into two relaxation channels the irreducible memory kernel $\mathcal{M}_\mu^{\alpha\beta}(q,t)$ assumes the form of a $2\times 2$ matrix. The diagonal elements, $\mathcal{M}^{\parallel\parallel}_{\mu}(q,t)$ and $\mathcal{M}^{\perp\perp}_\mu(q,t)$, represent the memory kernel parallel and perpendicular to the confinement respectively. Due to the underlying geometry the matrix is symmetric ($\mathcal{M}^{\parallel\perp}_\mu(q,t)=\mathcal{M}^{\perp\parallel}_\mu(q,t)$) with the nonvanishing off-diagonal elements describing the coupling between the two relaxation channels. This is a substantial difference to the diagonal approximation for the slit, where this coupling is discarded~\cite{Mandal:SoftMatter:2017,Jung:2020}. Since we are dealing with $2\times2$ matrices only, no approximations have to be invoked to solve the e.o.m.\ numerically.

\subsection{Effective memory kernel}
The set of coupled equations for the ISF, equation~\eref{eq:eom_correlator}, and the matrix-valued irreducible memory kernel, equation~\eref{eq:eom_irreducible}, completed with the closure relation~\eref{eq:MCT_functional} can be recast in a simplified form suitable for numerical integration by introducing an effective scalar memory kernel.

We use the convention
\begin{eqnarray}
 \hat{S}_{\mu}(q,z)=\rmi\int_0^\infty S_{\mu}(q,t)\exp(\rmi z t)\mathrm{d}t,  \quad \mbox{Im}[z]>0,
\end{eqnarray}
for the Fourier-Laplace transform with  complex frequency $z$ in the upper complex half plane $\mathbb{C}_{+}=\{z\in\mathbb{C}|\, \mbox{Im}[z]>0\}$. By linearity, the usual properties also transfer to the matrix-valued case. It readily follows that these are Nevanlinna functions with the following properties~\cite{Lang:JStatMech:2013}:
\begin{enumerate}
  \item[(1)] $\hat{S}_{\mu}(q,z)$ is analytic in $\mathbb{C}_{+}$.
  \item[(2)] $\hat{S}_{\mu}(q,-z^*)=-\hat{S}_{\mu}(q,z^*)$.
  \item[(3)] $\lim_{\eta\to\infty}\mbox{Im}[\hat{S}_{\mu}(q,z=\rmi\eta)]$ is finite.
  \item[(4)] $\mbox{Im}[\hat{S}_{\mu}(q,z)] \ge  0$ for $z\in\mathbb{C}_{+}$.
\end{enumerate}

Then both e.o.m., equations~\eref{eq:eom_correlator} and~\eref{eq:eom_irreducible} can be rewritten in the Laplace domain
\begin{eqnarray}\label{eq:eom_laplace}
\left[ z + \rmi D_\mu(q) S_\mu(q)^{-1} + \delta \hat{K}_\mu(q,z) S_\mu(q)^{-1} \right]\hat{S}_{\mu}(q,z) = - S_\mu(q),\\
\left[\rmi\boldsymbol{\mathcal{D}}_{\mu}(q)^{-1}+\hat{\boldsymbol{\mathcal{M}}}_{\mu}(q,z)\right] \delta\hat{\boldsymbol{\mathcal{K}}}_{\mu}(q,z)=-\rmi\hat{\boldsymbol{\mathcal{M}}}_{\mu}(q,z)\boldsymbol{\mathcal{D}}_{\mu}(q).
\end{eqnarray}
The second equation can be simplified by introducing 
$\hat{\boldsymbol{\mathcal{K}}}_{\mu}(q,z):=\delta\hat{\boldsymbol{\mathcal{K}}}_{\mu}(q,z)+\rmi\boldsymbol{\mathcal{D}}_{\mu}(q)$,
\begin{equation}
 \hat{\boldsymbol{\mathcal{K}}}_{\mu}(q,z)=-\left[\rmi\boldsymbol{\mathcal{D}}_{\mu}(q)^{-1}+\hat{\boldsymbol{\mathcal{M}}}_{\mu}(q,z)\right]^{-1}. 
\end{equation}
The cost of this simplification is that $\hat{K}_\mu(q,z) \to \rmi\boldsymbol{\mathcal{D}}_{\mu}(q)$ as $z\to\infty$ displays a nontrivial high-frequency limit. In particular, $\hat{K}_\mu(q,z)$ is not the Fourier-Laplace transform of a correlation function, rather it formally acquires an instantaneous relaxation via a temporal $\delta$-function.
The crucial step is now to introduce an effective memory kernel $M_\mu(q,t)$ implicitly defined via 
\begin{eqnarray}\label{eq:effective_kernel}
-\left[\rmi D_{\mu}(q)^{-1}+\hat{M}_{\mu}(q,z)\right]^{-1} := \hat{K}_{\mu}(q,z) =\rmi D_{\mu}(q)+\delta\hat{K}_{\mu}(q,z),
\end{eqnarray}
such that equation~\eref{eq:eom_laplace} becomes
\begin{eqnarray}\label{eq:eom_laplace_effective}
\fl\left[z +\rmi  D_\mu(q) S_\mu(q)^{-1} - \rmi z D_\mu(q) \hat{M}_\mu(q,z) \right] \hat{S}_\mu(q,z) =  - S_\mu(q) + \rmi D_\mu(q) \hat{M}_\mu(q,z) S_\mu(q), 
\end{eqnarray}
equivalent to an integro-differential equation in the time domain
\begin{eqnarray}\label{eq:eom_correlator_effective}
 \fl\dot{S}_{\mu}(q,t) + D_{\mu}(q)S_{\mu}(q)^{-1}S_{\mu}(q,t) + D_{\mu}(q)\int_0^t M_{\mu}(q,t-t^\prime)\dot{S}_{\mu}(q,t^\prime)\mathrm{d}t^\prime=0.
\end{eqnarray}

The equation for the effective memory kernel can be rearranged with the explicit matrix elements.  Suppressing the dependence on the complex frequency and the wavenumber for the moment, we arrive at
\begin{eqnarray}
 \fl\Big(\rmi D_{\mu}^{-1}+\hat{M}_{\mu}\Big)^{-1} = 
(-\rmi)\frac{q^2\left(D_0^{-1}-\rmi\hat{\mathcal{M}}_{\mu}^{\perp\perp}\right)+2q Q_{\mu}\rmi \hat{\mathcal{M}}_{\mu}^{\parallel\perp}+Q_{\mu}^2\left(D_0^{-1}-\rmi\hat{\mathcal{M}}_{\mu}^{\parallel\parallel}\right)}{\left(D_0^{-1}-\rmi\hat{\mathcal{M}}_{\mu}^{\parallel\parallel}\right)
  \left(D_0^{-1}-\rmi\hat{\mathcal{M}}_{\mu}^{\perp\perp}\right)+
  \left(\hat{\mathcal{M}}_{\mu}^{\parallel\perp}\right)^2}.
\end{eqnarray}
In the time domain the integral equation for the effective memory kernel becomes
\begin{eqnarray}\label{eq:effective_memory_time}
 D_{\mu}(q)  M_{\mu}(q,t) &+ D_0^2 \int_0^t M_{\mu}(q,t-t^\prime)\alpha_{\mu}(q,t^\prime)\mathrm{d}t^\prime  =
  D_0 \beta_{\mu}(q,t) \nonumber\\  &+ D_0^2\int_0^t \mathcal{M}_{\mu}^{\parallel\parallel}(q,t-t^\prime)
  \mathcal{M}_{\mu}^{\perp\perp}(q,t^\prime)\mathrm{d}t^\prime \nonumber \\ &- D_0^2\int_0^t \mathcal{M}_{\mu}^{\parallel\perp}(q,t-t^\prime)
  \mathcal{M}_{\mu}^{\parallel\perp}(q,t^\prime)\mathrm{d}t^\prime,
\end{eqnarray}
where
\begin{eqnarray}\label{eq:alpha_collective}
\fl\alpha_{\mu}(q,t) = Q_{\mu}^2\mathcal{M}_{\mu}^{\parallel\parallel}(q,t) + q^2\mathcal{M}_{\mu}^{\perp\perp}(q,t) - 2 q Q_{\mu}\mathcal{M}_{\mu}^{\parallel\perp}(q,t),
\end{eqnarray}
and
\begin{eqnarray}\label{eq:beta_collective}
\fl\beta_{\mu}(q,t) = \frac{q^2}{q^2+Q_{\mu}^2}
  \mathcal{M}_{\mu}^{\parallel\parallel}(q,t) + \frac{Q_{\mu}^2}{q^2+Q_{\mu}^2}\mathcal{M}_{\mu}^{\perp\perp}(q,t) + \frac{2q Q_{\mu}}{q^2+Q_{\mu}^2}\mathcal{M}_{\mu}^{\parallel\perp}(q,t),
\end{eqnarray}
only depend on the matrix elements of the matrix-valued memory kernel $\boldsymbol{\mathcal{M}}_\mu(q,t)$, but not on the scalar effective memory kernel $M_\mu(q,t)$.

\subsection{Mode-coupling approximation}
MCT approximates the  memory kernel in terms of a bilinear functional of the ISF itself. The functional for the quasi-confinement follows by direct translation from  
the slit case by taking only the diagonal elements
\begin{eqnarray}
 \fl\mathcal{M}_{\mu}^{\alpha\beta}(q,t) \approx  \frac{1}{2N}\sum_{\substack{\vec{q}_1 \cr \vec{q}_2=\vec{q}-\vec{q}_1}}
\sum_{\substack{\mu_1\cr \mu_2=\mu-\mu_1}} \mathcal{Y}_{\mu,\mu_1\mu_2}^\alpha (\vec{q};\vec{q}_1,\vec{q}_2) S_{\mu_1}(q_1,t) S_{\mu_2}(q_2,t) \mathcal{Y}_{\mu,\mu_1\mu_2}^\beta
 (\vec{q};\vec{q}_1,\vec{q}_2)^*,
\end{eqnarray}
where the coupling between different modes and wavenumbers is determined by the vertices
\begin{eqnarray}
\fl\mathcal{Y}_{\mu,\mu_1\mu_2}^\alpha  (\vec{q};\vec{q}_1,\vec{q}_2) = \frac{n_0}{L^2}\delta_{\vec{q},\vec{q}_1+\vec{q}_2}\delta_{\mu,\mu_1+\mu_2} \Big[b^\alpha(\hat{\vec{q}}\cdot\vec{q}_1,Q_{\mu_1})
  c_{\mu_1}(q_1) + (1 \leftrightarrow 2) \Big].
\end{eqnarray}
The vertices are uniquely determined by the equilibrium structure of the quasi-confined liquid with direct correlation function $c_\mu(q)$ which is related to the structure factor $S_\mu(q)$ by a generalized Ornstein-Zernike equation~\cite{Petersen:JStatMech:2019} $S_\mu(q) = 1/[1- n_0L^{-2} c_\mu(q)]$. Let us also  mention that 
in the long-wavelength limit the force kernel decouples  with respect to the channel index $\mathcal{M}_{\mu}^{\alpha\beta}(q\to 0,t)=: \delta^{\alpha\beta}\mathcal{M}_{\mu}^{\alpha}(t)$ and the vertex vanishes as $O(q)$~\cite{Schrack:PhilMag:2020}.

Taking the thermodynamic limit, sums over wave vectors are replaced by integrals
\begin{eqnarray}
\lim_{N\to \infty} \frac{1}{N}\sum_{\substack{\vec{q}_1\cr \vec{q}_2=\vec{q}-\vec{q}_1}}  \left( \ldots \right) \to \frac{1}{n_0}
  \int \frac{\mathrm{d} \vec{q}_1}{(2\pi)^2}  \left( \ldots \right),
\end{eqnarray}
where we abbreviate  $\vec{q}_2 = \vec{q}-\vec{q}_1$ throughout. 

We choose the $x$-axis along the  $\vec{q}$-direction and write $\vec{q}_1 = q_1 (\cos\vartheta, \sin\vartheta)$ with the polar angle $\vartheta$. Then the 2D integral simplifies considerably in bipolar coordinates
\begin{eqnarray}\label{eq:bipolar}
\int \frac{\mathrm{d} \vec{q}_1}{(2\pi)^2}  \left( \ldots \right) = \int \frac{q_1 \mathrm{d} q_1 \mathrm{d}\vartheta}{(2\pi)^2} \left( \ldots \right)  = \frac{1}{4\pi^2} \int_0^\infty \mathrm{d}q_1\int_{|q-q_1|}^{q+q_1} \frac{2q_2\mathrm{d}q_2}{q \left|\sin\vartheta\right|}  \left( \ldots \right),
\end{eqnarray}
where we used a change of variable from the angle $\vartheta$ to $q_2$ via the cosine law $q^2 + q_1^2 - 2 q q_1 \cos\vartheta = q_2^2$. Then for the sine one readily calculates
\begin{equation}
 \left|\sin\vartheta\right|=
 \frac{\sqrt{4 q_1^2 q^2 - (q_1^2 + q^2 - q_2^2)^2}}{2q q_1} . 
\end{equation}
The factor 2 in equation~\eref{eq:bipolar} arises due to a split of the integral into the two half planes, where the bipolar coordinates are uniquely determined, $0\leq \vartheta < \pi$ and $\pi \leq  \vartheta < 2 \pi$.

The explicit expression for the memory kernel is then given by
\begin{eqnarray}\label{eq:memory_MCT}
 \mathcal{M}_\mu^{\alpha\beta}(q,t) =& \frac{3 \varphi}{4L^3\pi^3\sigma^3} \int_0^\infty  \mathrm{d}q_1\int_{|q-q_1|}^{q+q_1} \frac{q_2 \mathrm{d} q_2 }
  {\sin\vartheta} \sum_{\substack{\mu_1\cr \mu_2=\mu-\mu_1}} S_{\mu_1}(q_1,t) 
  S_{\mu_2}(q_2,t) \nonumber\\ & \times\left[b^\alpha\left(\hat{\vec{q}}\cdot \vec{q}_1,Q_{\mu_1}\right)c_{\mu_1}(q_1)+\left(1\leftrightarrow 2\right)\right]   \nonumber\\ &\times\left[b^\beta\left(\hat{\vec{q}}\cdot \vec{q}_1,Q_{\mu_1}\right)c_{\mu_1}(q_1)+\left(1\leftrightarrow 2\right)\right],
\end{eqnarray}
with $\hat{\vec{q}}\cdot \vec{q}_1 = q_1 \cos \vartheta =  (q^2 + q_1^2 - q_2^2 )/2 q$.

\subsection{Glass transition}\label{sec:nonergodicity_theory}
Within MCT glassy states are characterized by nonvanishing long-time limits of the wavenumber-dependent ISF
\begin{equation}
 F_{\mu}(q) :=\lim_{t\to\infty}S_{\mu}(q,t)=-\lim_{z\to 0} z \hat{S}_{\mu}(q,z).
\end{equation}
These nonergodicity parameters $F_\mu(q)$ are directly accessible in simulations or experiments and encode valuable information about the arrested structure of the confined liquid. 
In contrast to the glassy state, for liquid states the nonergodicity parameters evaluate to zero. \Eref{eq:eom_laplace_effective} yields in the low-frequency limit a relation for the nonergodicity parameter
\begin{equation}\label{eq:nonerg}
 F_\mu(q) = S_\mu(q) - \left[ S_\mu(q)^{-1} + N_\mu(q) \right]^{-1}, 
\end{equation}
with the long-time limit
\begin{equation}
 N_{\mu}(q):=\lim_{t\to\infty}M_{\mu}(q,t) =-\lim_{z\to 0}z \hat{M}_{\mu}(q,z),
\end{equation}
of the effective memory kernel. The relation to the long-time limit of the irreducible memory kernel 
\begin{equation}
 \boldsymbol{\mathcal{N}}_{\mu}(q):=\lim_{t\to\infty}\boldsymbol{\mathcal{M}}_{\mu}(q,t) =
  -\lim_{z\to 0}z \hat{\boldsymbol{\mathcal{M}}}_{\mu}(q,z),
\end{equation}
is similar to equation~\eref{eq:effective_kernel}
\begin{eqnarray}\label{eq:effective_memory_nonerg}
N_{\mu}^{-1}=\vec{k}\cdot \boldsymbol{\mathcal{N}}_{\mu}^{-1}\cdot\vec{k} = \frac{q^2\mathcal{N}_{\mu}^{\perp\perp}-2 q Q_{\mu}\mathcal{N}_{\mu}^{\parallel\perp}+Q_{\mu}^2\mathcal{N}_{\mu}^{\parallel\parallel}}{\mathcal{N}_{\mu}^{\parallel\parallel}\mathcal{N}_{\mu}^{\perp\perp}-\left(\mathcal{N}_{\mu}^{\parallel\perp}\right)^2},
\end{eqnarray}
where a compact notation $\vec{k}=(\vec{q},Q_\mu)$ has been introduced and  the wavenumber dependence $q$ has been suppressed. It can be seen that the set of self-consistent equations for the ISF can be solved for their long-time limits without solving explicitly for the full dynamics~\cite{Lang:PRE:2012}. In general, the set of equations~\eref{eq:nonerg} to~\eref{eq:effective_memory_nonerg} has many solutions, in particular there is always the trivial solution $F_{\mu}(q)=0$. The concept of an effective memory kernel ensures that the covariance and maximum principle are fulfilled for \mbox{(quasi-)}confined liquids. Then the nonnegative solution for the long-time limit of the ISF is maximal and can be calculated by iterating the fixed-point equation without solving for the full dynamics explicitly~\cite{Lang:PRE:2012,Lang:JStatMech:2013}.
The MCT approximation, equation~\eref{eq:memory_MCT}, for the nonergodicity parameter is simply obtained by replacing the ISF $S_{\mu}(q,t)$ with the corresponding long-time limits $F_{\mu}(q)$. To determine the critical point, equation~\eref{eq:nonerg} is solved iteratively using $F_\mu^{(0)}(q) = S_\mu(q)$ as the starting value.

\subsection{Numerical implementation}
We investigate a quasi-confined fluid with hard spheres of diameter $\sigma$ and bare  diffusion coefficient $D_0$. Then, $\sigma$ sets the unit of length and $\sigma^2/D_0$ the unit of time. The wavenumbers are discretized on a uniform grid $q=\hat{q}\Delta q + q_0, \hat{q} = 0, \ldots, N_q-1$ parallel to the confinement with parameters $q_0\sigma=0.1212$, $\Delta q\, \sigma=0.4040$ using $N_q=100$ grid points. These optimized values ensure that the numerical solution can be obtained efficiently with a sufficiently accurate resolution. The discrete mode indices related to the confining direction are truncated to $|\mu|\leq 15$. The algorithm for obtaining the time-dependent memory kernels and correlators is described in~\ref{sec:appendix_numerics}.

\section{Simulations}\label{sec:simulation}
We perform event-driven simulations of hard spheres undergoing Newtonian dynamics in 3D~\cite{Rapaport:2004}. Periodic boundary conditions are employed, such that the length of the $z$-dimension of the simulation box, $L$, is much smaller than the other two dimensions. The periodic box in the $x$- and $y$-directions has a length $50.0\sigma$. While MCT considers monodisperse particles, it is necessary to add polydispersity to the simulations to avoid crystallization. We use an inverse-occupied volume distribution of particle sizes, as this distribution has been shown to effectively suppress crystallization~\cite{Ninarello:PRX:2017}. The probability density for the particle diameter $s$ is given by  $P(s)\propto 1/s^3$, $s\in \left[\sigma_{\mbox{\scriptsize min}},\sigma_{\mbox{\scriptsize max}}\right]$, where $\sigma_{\mbox{\scriptsize max}}$ is set to $1.25\sigma$ and $\sigma_{\mbox{\scriptsize min}}$ is chosen such that the mean particle diameter is $\sigma$. The standard deviation of this distribution is $0.117\sigma$, corresponding to a polydispersity of $11.7\%$.
To initialize the polydisperse system with a high packing fraction in an allowed configuration where no particles are overlapping, we use the standard compression algorithm~\cite{Woodcock:Annals:1981, Li:EPL:2008}.
The particles are initialized on a cubic lattice, with reduced particle size such that the packing fraction is 0.1. As the particles move, all particle radii are increased at a rate of $0.01 \sigma/t_0$ until the desired packing fraction is reached. The system is then run for a further sample-preparation time of  $10^3 t_0$ using standard event-driven dynamics before measurements are taken. 
The thermal energy $k_BT$ and the particle mass $m$ enter only via the time scale of the simulation $t_0=\sqrt{m\sigma^2/k_BT}$.
The dynamics in the system are quantified by calculating the collective ISF, equation~\eref{eq:ISF}, from the density modes, equation~\eref{eq:denmodes}. To ensure good statistics with reasonable memory requirements for long simulations, the configuration is sampled with an order-N algorithm~\cite{frenkel2001understanding}. The data is collected over a time $10^4 t_0$, and 200 independent copies of each simulation are used to improve statistics. While the simulations use Newtonian dynamics, we expect our results for the long-time dynamic properties to be similar to those for Brownian dynamics~\cite{Franosch:JNCS:1998,Hunter:RepProgPhys:2012,Pusey:PhilTransRoyal:2009}, therefore it is reasonable to compare them to the MCT results.

Recent simulations relying on advanced algorithms~\cite{Berthier:PRL:2016,Ninarello:PRX:2017,Lindquist:JCP:2018, Bommineni:PRL:2019}
have revealed that polydispersity does not suppress crystallization completely and eventually the system demixes. Therefore our simulations are only in a metastable state which is subject to aging effects. Increasing the sample preparation  time before starting the measurements by a factor of 20 leads to a slight drift in the data, however, the trends we identify below and the resulting conclusions are not affected. Details on the issue of demixing and the extended simulation results are presented in \ref{sec:appendix_ISF}.

\section{Results and discussions}\label{sec:result}
In this section we  investigate the ISF $S_\mu(q,t)$ as well as the nonergodicity parameters $F_\mu(q)$ comparing results from solving the MCT equations numerically with those from simulations. 
For completeness, we also show the static structure factors. Finally, we provide a nonequilibrium-state diagram within the framework of MCT.

\subsection{Static structure factor}
As input for MCT we use the static quantities of quasi-confined liquids elaborated in Ref.~\cite{Petersen:JStatMech:2019}. Therefore we start our discussion with a short comparison between the static structure factors for liquid state theory using Percus-Yevick (PY) closure relation and event-driven simulations. Since we are interested in the behavior close to the glass transition, relatively high packing fractions are considered which require polydisperse systems in simulations, contrary to the aforementioned work.

\begin{figure}[h]
\begin{minipage}[b]{\linewidth}
\centering
\includegraphics{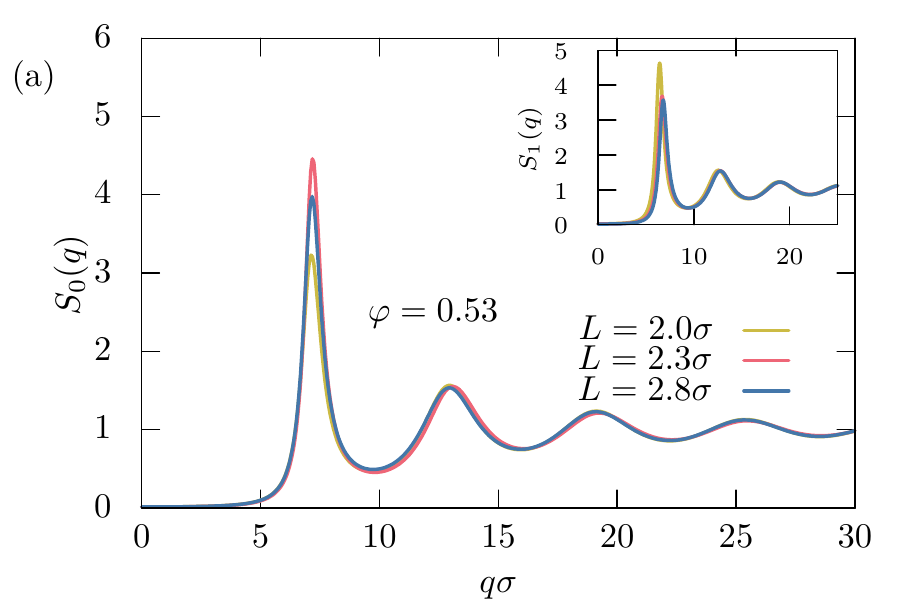}
\end{minipage}
\begin{minipage}[b]{\linewidth}
\centering
\includegraphics{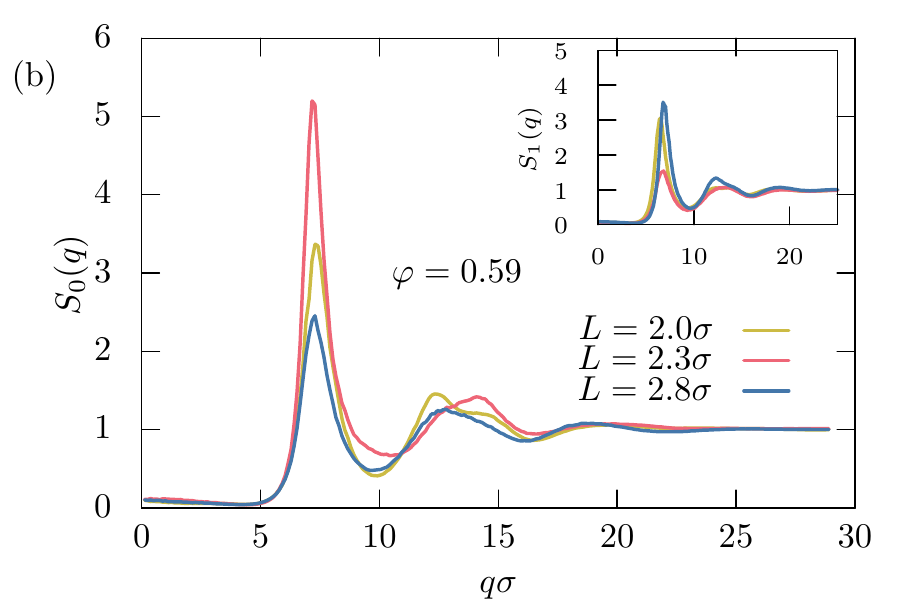}
\end{minipage}
\caption{Static structure factors $S_0(q)$ and $S_1(q)$ (inset) as a function of the wavenumber $q$ for different values of $L$ at packing fraction $\varphi=0.53$ for MCT (a) and $\varphi=0.59$ for simulations (b). }
\label{fig:structure}
\end{figure}

It is known from bulk systems that MCT underestimates the critical packing fraction for the glass transition by approximately $20\%$. Therefore the static structure factors for different confinement lengths $L$ within the framework of liquid state theory at packing fraction $\varphi=0.53$ [\Fref{fig:structure}(a)] are compared with simulation data at a higher packing fraction of $\varphi=0.59$ [\Fref{fig:structure}(b)].

The in-plane structure factor $S_0(q)$ sensitively depends on the confinement length $L$ for both theory and simulations, which becomes particularly apparent in the height of the first peak. As already discussed in Ref.~\cite{Petersen:JStatMech:2019} the curve shapes are similar but there are quantitative differences between simulations and the structure factors calculated with PY closure relation. Additionally, due to polydispersity the second (and subsequent) peaks are washed out for simulations. The nonmonotonic behavior of the height of the first structure factor peak measuring the near-ordering of the fluid indicates that the the ordering changes from commensurable to incommensurable packing. The particle diameter $\sigma$ and the confinement length $L$ are said to be commensurable if $L/\sigma$ is an integer number and incommensurable if $L/\sigma$ is a half-integer. This interpretation corroborates that the most incommensurable length $L=2.3\sigma$ shows the highest degree of ordering.

Differences are also present in higher order modes, e.g.\ $S_1(q)$ (inset), taking into account the arrangement along the confining direction, but these modes are less important for the dynamic quantities as will be discussed below.

In summary, the Percus-Yevick closure yields static structure factors that qualitatively agree with the simulations, in particular, it reproduces the trends upon varying the confinement length. Since we are 
interested only in predicting trends for the dynamics, we rely on the PY structure factors for the dynamic MCT calculations rather than the measured ones from simulations.    

\subsection{Intermediate scattering function}
We proceed with the dynamical behavior of  the ISF close to the glass transition for a  confinement length of  $L=2.8\sigma$  for which in the theory the static properties display neither strong commensurable nor incommensurable packing. 

\begin{figure}[t]
 \centering
  \includegraphics{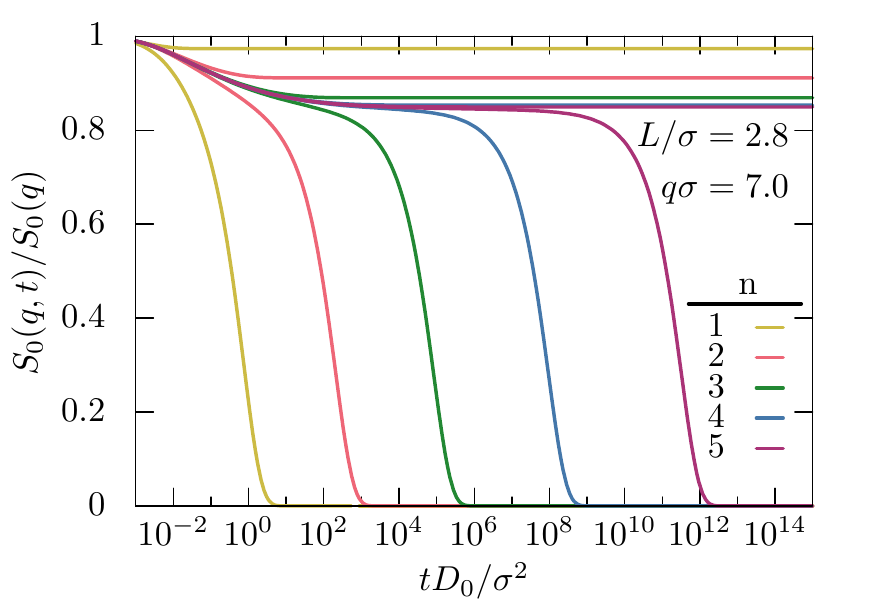}
  \caption{Normalized ISF $S_0(q,t)/S_0(q)$ for different $\epsilon=(\varphi-\varphi_c)/\varphi_c=\pm 10^{-n}$ at wavenumber $q\sigma= 7.0$ and confinement length $L=2.8\sigma$. Liquid curves ($\epsilon< 0$) approach the glass transition from left to right, corresponding glass curves ($\epsilon> 0$) from top to bottom.}
  \label{fig:numerics_eps}
\end{figure}

\Fref{fig:numerics_eps} shows the normalized ISF $S_0(q,t)/S_0(q)$ for mode index $\mu=0$ close to the critical point $\varphi_c$,  for several separation parameters $\epsilon=(\varphi-\varphi_c)/\varphi_c$. 
The wavenumber $q\sigma= 7.0$ corresponds to  the first structure factor peak in $S_{0}(q)$. Just as in the case of a bulk liquid there are two possibilities for the dynamic evolution of the ISF. After an initial decay present at all packing fractions, for $\epsilon>0$ the theoretical results obtained from MCT converge to a nonzero value characterizing the glassy state. 
The structure cannot completely relax and the nonvanishing long-time limit of the normalized ISF then corresponds to the normalized nonergodicity parameter $F_\mu(q)/S_\mu(q)$. Alternatively, for $\epsilon<0$ a two-step relaxation with an extended intermediate plateau is clearly visible. The plateau expands for decreasing $|\epsilon|$ and persists over several orders of magnitude in time for small $|\epsilon|$. At the critical point $\varphi_c$ for $\epsilon\to 0$ the structural relaxation time diverges.

\begin{figure}[t]
\begin{minipage}[b]{\linewidth}
\centering
\includegraphics{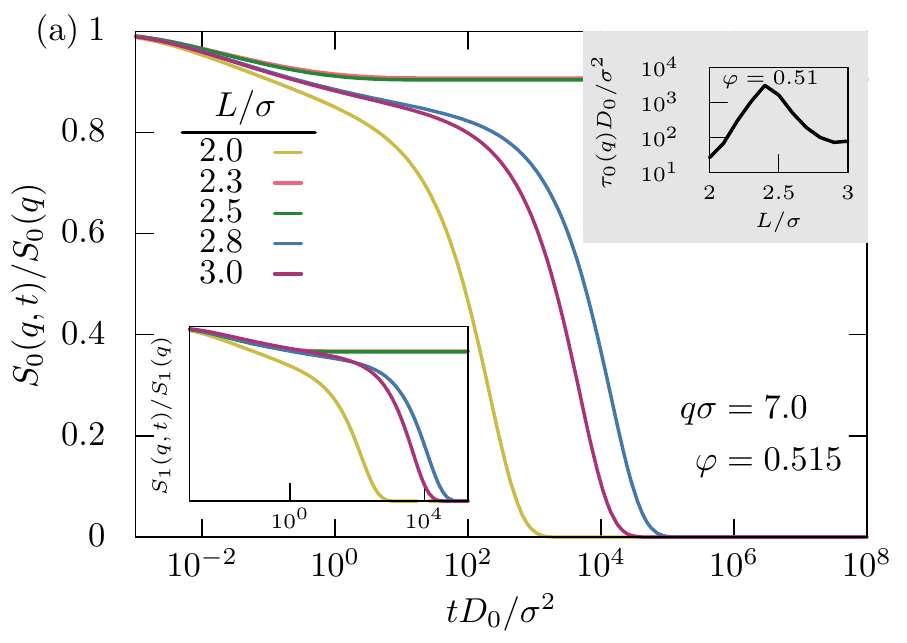}
\end{minipage}
\begin{minipage}[b]{\linewidth}
\centering
\includegraphics{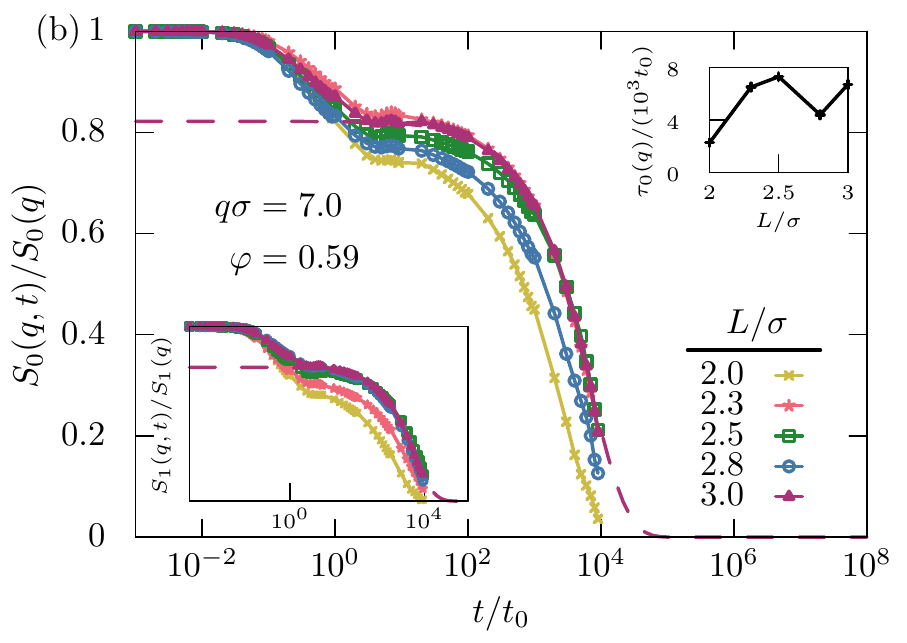}
\end{minipage}
\caption{Normalized ISF $S_0(q,t)/S_0(q)$ and $S_1(q,t)/S_1(q)$ (lower left insets) for $q\sigma=7.0$ at (a) packing fraction $\varphi=0.515$ for MCT and (b) $\varphi=0.59$ for simulations, respectively.
  The dashed line in (b) is a  KWW fit for the largest confinement length. The upper right insets show the  relaxation time $\tau_0(q)$  as a function of the confinement length for  (a)  MCT   (at $\varphi=0.51$)  on logarithmic scales and (b)  simulations on linear scales.
 }
\label{fig:ISF}
\end{figure}

We want to analyze how the quasi-confinement affects the dynamics of our system. \Fref{fig:ISF} presents the temporal evolution of the normalized ISF for 
different confinement lengths $L$ for the wavenumber $q\sigma= 7.0$ at constant packing fraction, comparing MCT results for $\varphi=0.515$ [\Fref{fig:ISF}(a)] with simulation data at a higher packing fraction of $\varphi=0.59$ [\Fref{fig:ISF}(b)].  The discussion is restricted to the dynamics of the first two modes, $S_0(q,t)/S_0(q)$ and $S_1(q,t)/S_1(q)$ [lower inset \Fref{fig:ISF}].

Within MCT after the initial decay, independent of the confinement length, at intermediate times either a glassy plateau ($L=2.3\sigma$) manifests itself or a stretched relaxation ($L=2.0\sigma$ and $L=2.8\sigma$) is observed. Thus, the system is in a liquid state in the case of commensurate packing. By increasing the confinement length to a more incommensurate value $L=2.3\sigma$, the plateau indicates the transition to a glassy state. The reason for this nonmonotonic behavior is attributed to the competition between the local ordering of the hard spheres and confinement effects. If the confinement length is an integer multiple of the particle diameter, it allows for large longitudinal diffusion, whereas the dynamics are slowed down in case of incommensurate packing. The structural relaxation is about one order of magnitude faster for $L=2.0\sigma$ compared to $L=2.8\sigma$.
The nonmonotonic behavior of the dynamical solution for $S_0(q,t)/S_0(q)$ and $S_1(q,t)/S_1(q)$ show no significant differences. This is remarkable since $S_0(q,t)$ only considers the dynamics in the lateral direction in contrast to higher order modes of the ISF which also include the dynamics in the confining direction. These significant variations between different modes are present for example in the static structure factors $S_0(q)$ and $S_1(q)$, cf.~\Fref{fig:structure}. We conclude that $S_0(q)$, in particular the first sharp diffraction peak, is the relevant quantity also for the dynamics of $S_1(q,t)$. 

In contrast, the simulation data for $S_0(q,t)/S_0(q)$ show only an intermediate plateau for all investigated confinement lengths, followed by a stretched  relaxation. The height of the plateau depends nonmonotonically on $L$ just as for the MCT solution, beside $L=3.0$. The structural relaxation is about half an order of magnitude faster for commensurate packing ($L=2.0\sigma$) compared to more incommensurate values ($L=2.3\sigma$). In contrast to the MCT solution, the qualitative behavior differs between $S_0(q,t)/S_0(q)$ and $S_1(q,t)/S_1(q)$ in the simulations, where the plateau values increase monotonically upon enlarging the confinement length. 
Furthermore, no ideal glass transition occurs in the simulations. The overall dependence of the structural relaxation on the confinement length in simulations is less drastic than in MCT, which is probably due to the polydispersity. Then the effects of commensurate and incommensurate packing should be less pronounced, which coincides with our observation. Nevertheless, the prominent nonmonotonic behavior is present in both theory and simulations.

The simulation data for the ISF in~\Fref{fig:ISF} are fit to the phenomenological Kohlrausch-William-Watts (KWW) stretched exponential~\cite{Williams:Faraday:1970}
\begin{equation}\label{eq:KWW}
\mbox{KKW fit: } 
 S_\mu(q,t)=F_\mu(q)\exp\left\{-\left[t/\tau_\mu(q)\right]^{\beta_\mu(q)} \right\},
\end{equation}
with the Kohlrausch exponent $\beta_\mu(q)$ and the relaxation time $\tau_\mu(q)$ in the range $t\in(10t_0,10^4t_0)$. Similarly, we identify the relaxation time $\tau_\mu(q)$ within MCT with the time where the ISF has reduced to $1/e$. The nonmonotonic dependence of $\tau_\mu(q)$ on the confinement length is shown in the upper insets of \Fref{fig:ISF}. To ensure that all correlators decay to zero we compare the simulation results with MCT results for a slightly lower packing fraction $\varphi=0.51$. Qualitatively, the behavior is quite similar, with the maximum relaxation time occurring at almost the same length ($L=2.4$ within MCT and $L=2.5$ for simulations, respectively). There are deviations for large $L$ probably related to the static input where differences are most pronounced for large confinement lengths.

It has been shown recently~\cite{Jung:JStatMech:2020} that a well defined $\beta$-scaling equation is valid for confined liquids with multiple relaxation channels using only moderate assumptions. These assumptions are all fulfilled in the framework of quasi-confinement, therefore in principle a full asymptotic analysis of the dynamics is possible.

\subsection{Nonergodicity parameters from MCT and simulations}\label{subsec:nonergodicity}

\begin{figure}[h]
\begin{minipage}[b]{\linewidth}
\centering
\includegraphics{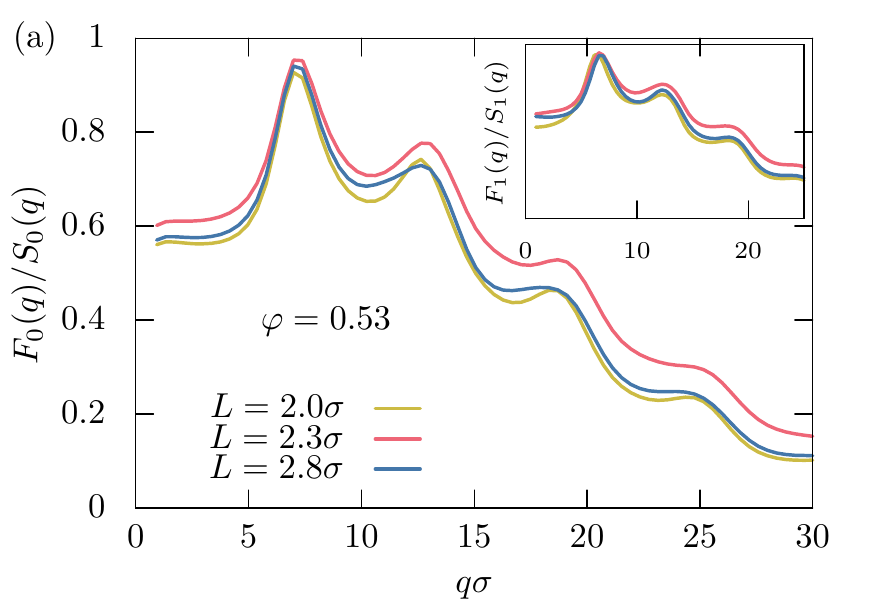}
\end{minipage}
\begin{minipage}[b]{\linewidth}
\centering
\includegraphics{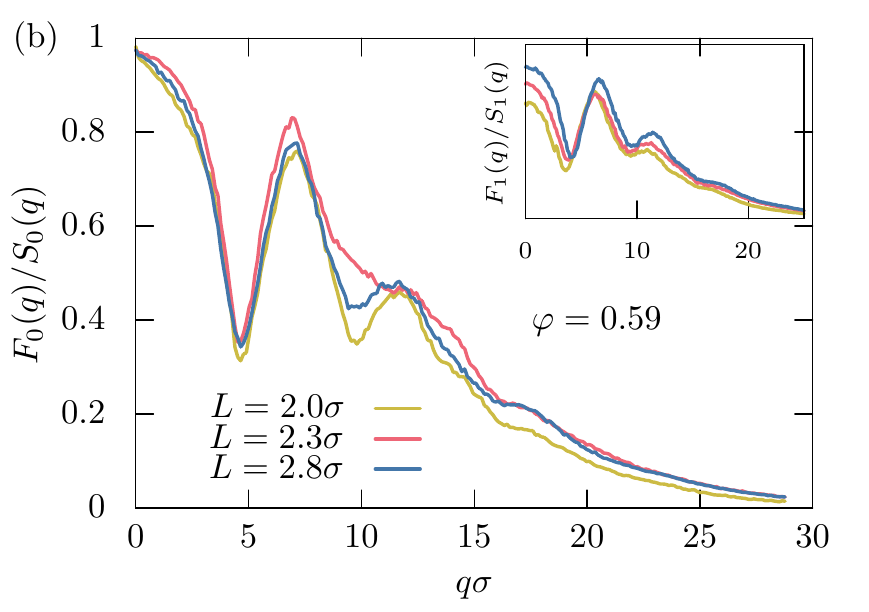}
\end{minipage}
\caption{Normalized nonergodicity parameters $F_0(q)/S_0(q)$ and $F_1(q)/S_1(q)$ (inset) as a function of the wavenumber $q$ for different values of $L$ at packing fraction $\varphi=0.53$ for MCT (a) and $\varphi=0.59$ for simulations (b).}
\label{fig:nonergodicity}
\end{figure}

In this subsection we analyze the nonergodicity parameters of the quasi-confined hard-sphere system. The numerical iteration to calculate these parameters within MCT has been described in~\sref{sec:nonergodicity_theory}. 
The nonergodicity parameters from the simulations are identified with the plateau value of the KWW stretched exponential,~\Eref{eq:KWW}, which is fit to the simulation data for all values of $q$.

We present results for confinement lengths $L$ where the effects are most pronounced in~\Fref{fig:nonergodicity} for the first two modes $F_0(q)/S_0(q)$ and $F_1(q)/S_1(q)$ both for MCT  and simulations. The wavenumber dependence is similar to the oscillations of the related static quantities and the variations with the confinement length reflect the nonmonotonic evolution of the static structure factor, \Fref{fig:structure}. The qualitative behavior is also quite similar to a liquid confined between two parallel hard walls~\cite{Mandal:SoftMatter:2017}, although the nonmonotonic effects are less pronounced due to the absence of layering.

The long-wavelength limit differs significantly between theory and simulations. This feature is also present in the slit case~\cite{Mandal:SoftMatter:2017} and in bulk liquids, where it is rationalized by the effect of polydispersity~\cite{Weysser:PRE:2010}. 
The height of the first peak of the nonergodicity parameter also differs between theory and simulations, which we attribute to differences in the static structure factor peak, \Fref{fig:structure}, serving as  only input to the MCT equations. However, the match seems slightly better than for the slit case~\cite{Mandal:SoftMatter:2017}.

For the first higher mode $F_1(q)/S_1(q)$ the wavenumber dependence is qualitatively  similar to $F_0(q)/S_0(q)$. This is remarkable since   the static structure factors $S_1(q)$ are significantly different from $S_0(q)$, 
especially their dependence on the confinement length, \Fref{fig:structure}. Therefore, we  corroborate that the in-plane structure $S_0(q)$ is the relevant determinant for the particle dynamics even for the higher modes  just as in the slit case~\cite{Mandal:SoftMatter:2017}. Whereas the nonmonotonic dependence on the confinement length  for $F_1(q)/S_1(q)$ resembles the one of  $F_0(q)/S_0(q)$ in the MCT numerics, it differs
 within simulations. For example, the highest structural arrest in $F_1(q)/S_1(q)$ is reached for $L=2.8\sigma$ which is clearly visible in the first peak of the nonergodicity parameter.

\subsection{Nonequilibrium-state diagram}
\begin{figure}[h]
\centering
  \includegraphics{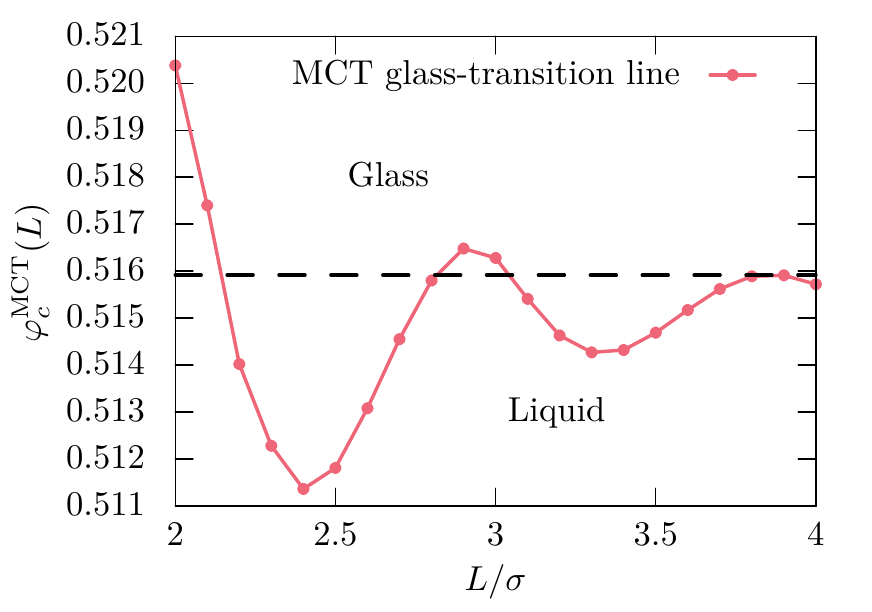}
\caption{Nonequilibrium-state diagram from MCT for a quasi-confined hard-sphere fluid. The dashed-line indicates the critical packing fraction for a 3D bulk  hard-sphere system.}
\label{fig:numerics_phasediagram}
\end{figure}

We use a simple bisection method to determine the glass-transition line as a function of the two control parameters $\varphi$ and $L$. Starting with two packing fractions, one within the liquid, $\varphi_l$, with vanishing nonergodicity parameter $F_\mu(q)=0$, and one in the glassy regime, $\varphi_g$, with finite $F_\mu(q)\ne 0$, the nonergodicity parameter for the intermediate packing fraction $\varphi_m=(\varphi_l+\varphi_g)/2$ is calculated. If it vanishes $\varphi_m$ is taken as the new reference point for the liquid state, otherwise it replaces $\varphi_g$. The procedure is continued until an accuracy of $10^{-5}$ is reached. Using this algorithm we can calculate the critical packing fraction as a function of the confinement length.  The glass-transition line found with this method defines a transition between ergodic liquid-like states and nonergodic glassy states, and as such we refer to \Fref{fig:numerics_phasediagram} as a nonequilibrium-state diagram.

We observe a nonmonotonic behavior of the critical packing fraction with an oscillation period comparable to the hard-sphere diameter $\sigma$. These oscillations are quite similar to the behavior for a system confined within a slit~\cite{Mandal:NatComm:2014}, however, the amplitude is approximately one order of magnitude smaller. Nevertheless, a reentrant behavior on lines with constant packing fraction can still be observed notwithstanding that the density profile of our fluid is constant. Therefore we conclude that the nonmonotonic behavior in slit geometry is not only a reflection of the layering within the fluid. The local packing is also significantly affected by the confinement and this leads to similar trends in the static structure factors and the nonequilibrium-state diagram. 

In our system the critical packing fraction roughly oscillates around the critical packing fraction for a 3D bulk system~\cite{Franosch:PRE_55_6:1997}, $\varphi_c^{3D} \approx 0.516$,
which is almost reached in the limit of large confinement length, $L\gtrsim 4\sigma$. The highest value of the critical packing fraction $\varphi_c\approx 0.520$ at $L=2.0\sigma$ differs significantly from the value predicted by MCT for hard disks~\cite{Bayer:PRE:2007}, $\varphi_c^{2D} \approx 0.697$. Therefore, we conclude that in our quasi-confined liquid extreme confinement cannot be considered as a two dimensional system, since the hard spheres are still able to move perpendicular to the confinement due to the periodic boundary conditions.

It remains a challenge to test the MCT prediction in simulations.  First of all, the nonmonotonic behavior of the glass-transition line in the framework of MCT only covers a small range of packing fractions. Together with the fact that there is no ideal 
glass transition in the simulations of hard spheres (the structure will eventually fully relax) and the polydispersity of the hard-sphere system,  small nonmonotonic effects within the nonequilibrium-state diagram would be hardly observable. However, the nonmonotonic effects shown in the state diagram are clearly reflected in the nonergodicity parameters and the ISF in the simulations.

\section{Summary and conclusions}\label{sec:conclusion}
In this work we have studied the collective dynamics of quasi-confined hard-sphere liquids by MCT and event-driven simulations. Not only are these quasi-confined liquids conceptually interesting by themselves, but they also constitute a valuable intermediate step between bulk liquids and liquids in confinement with boundaries. We have elaborated MCT equations accounting for relaxation channels parallel and perpendicular to the confinement direction. 
A stable numerical algorithm to generate solutions of the modified MCT equations has been suggested by introducing an effective memory kernel encoding the multiple channel relaxation. Numerical solutions of the MCT dynamics for the ISF can be generated up to arbitrarily long times.

The main observation is a nonmonotonic behavior on the confinement length for both theory and simulations, which becomes apparent for instance in the nonergodicity parameters. Furthermore, we have extracted a 
nonequilibrium-state diagram exhibiting a reentrant glass transition at paths of constant packing fraction. 

The nonmonotonic behavior can be interpreted by means of the local packing of the hard spheres. Then, the corresponding dynamics between commensurate and incommensurate packing change from a sliding motion to a more obstructed movement. Investigating not only the lowest mode, which only considers the parallel dynamics, but also higher order modes of the ISF we have concluded that the in-plane structure dominates the dynamics. Although results from MCT and simulations coincide quite well 
qualitatively and in particular, trends are correctly identified, no quantitative agreement was  achieved. We attribute the discrepancies to the polydispersity of the simulations, the errors in the static structure factors as provided by integral equation theory, and the well-known shortcomings  of  MCT  already present in bulk systems.

In contrast to a slit geometry, the density profile for quasi-confined liquids is uniform. Therefore, we are able to disentangle the effects of layering and confinement. We have shown that nonmonotonic behavior arises purely from the interplay of confinement and local order and corroborate that it is not simply a reflection of layering due to an inhomogeneous density profile. Nevertheless, the effects are less pronounced compared to the slit geometry due to the absence of layering. Therefore MCT provides the microscopic foundation for the empirical observations~\cite{Mittal:PRL:2006, Mittal:PRL:2008, Goel:PRL:2008, Ingebrigsten:PRL:2013, Bollinger:JCP:2015, Ingebrigtsen:PNAS:2018} that local quantities such as the density or the excess entropy play the dominant role for transport properties in confined liquids. We expect that this insight will also be relevant to  experiments\cite{Nugent:PRL:2007, Edmond:PRE:2012, Sarangapani:PRE:2011, Sarangapani:SoftMatter:2012, Eral:PRE:2009, Eral:Langmuir:2011,  Nygard:PRL:2012, Nygard:JCP:2013, Nygard:PRX:2016, Nygard:PRL:2016, Nygard:PCCP:2017, Nygard:PRE:2017}, where the effects of layering and confinement cannot be disentangled so easily.

We emphasize that for quasi-confined liquids translational symmetry in the confining direction ensures that no additional approximations are necessary. Furthermore, a coupling between the parallel and perpendicular relaxation channel naturally occurs in contrast to the full diagonal approximation for the slit geometry~\cite{Lang:PRL:2010,Lang:PRE:2012}, where the different relaxation channels are only linked by the MCT functional.

Our results motivate us to further investigate quasi-confined liquids. The MCT approach can be extended to the tagged-particle dynamics including the self-intermediate scattering function, the mean-square displacement or the velocity-autocorrelation function. These quantities will contribute to our physical understanding of (quasi-)confined liquids and are more easily accessible in computer simulations. 

The implications of our work are not limited to simple hard-sphere systems  but are also relevant for more complex systems. 
Additionally, we expect that a similar strategy to ours could be used to extend modified MCT approaches, such as systems of active Brownian particles (ABP)~\cite{Farage:PRE:2015, Liluashvili:PRE:2017, Szamel:JCP:2019} or driven granular spheres~\cite{Kranz:PRL:2010,Sperl:EPL:2012,Kranz:PRE:2013}, to quasi-confined geometries.

\ack
We thank Markus Gruber and Matthias Fuchs for useful discussions. This work has been supported  by the Austrian Science Fund (FWF): I 2887. CFP gratefully acknowledges a Lise-Meitner fellowship of the Austrian Science Fund (FWF): M 2471. The computational results presented have been achieved in part using the HPC infrastructure LEO of the University of Innsbruck.

\appendix
\section{Numerical solution for the time-dependent quantities}\label{sec:appendix_numerics}
We rely on a decimated time grid to cover several orders of magnitude in time~\cite{Fuchs:JoP:1991}
\begin{eqnarray}
 t_i=ih2^d,
\end{eqnarray}
for the numerical solution of time-dependent quantities with fundamental time step $h$, time point $0\le i \le N$ and decimation level $0\le d\le D$. For our numerical results we use $N=256$, $D=100$ and $h=10^{-9}\sigma^2/D_0$. In the following we abbreviate $i\equiv t_i$ for the discretized time grid. 

For $d=0$ the time-dependent quantity $\Phi(i)$ is initialized by an appropriate short-time solution. The procedure described as decimation maps the already known solution $\Phi(i)$ at decimation level $d-1$ to the first $N/2$ points of the coarser time grid at decimation level $d$
\begin{eqnarray}
 \Phi^{(d)}(i)=\Phi^{(d-1)}(2i), \quad i=1,\dots,N/2.
\end{eqnarray}
The solution for $i\ge N/2$ is calculated iteratively by discretizing the corresponding integral (integro-differential) equation as discussed below.

For the numerical solution of the effective memory kernel we transform the integral equation to an integro-differential equation. Then, the solution procedure is valid for both the effective memory kernel as well as for the correlator. The underlying basic algorithm has been described in appendices C and D of Ref.~\cite{Gruber:PRE:2016}. A more detailed description discussing the benefits of using an integro-differential method can be found in \cite{Gruber:PHD:2019}.

The integral equation for the effective memory kernel equation~\eref{eq:effective_memory_time} has the form of a Volterra integral equation
\begin{eqnarray}
 \Phi(t)+\int_0^t K(t-t^\prime)\Phi(t^\prime)\mathrm{d} t^\prime = f(t),
\end{eqnarray}
for each wavenumber $q$ and mode index $\mu$ with a given function $f(t)$ and kernel $K(t)$. In our case $\Phi(t)$ is proportional to $M_\mu(q,t)$, $K(t)$ to $\alpha_{\mu}(q,t)$ and $f(t)$ corresponds to the whole right-hand side of equation~\eref{eq:effective_memory_time}.

Differentiating the equation with respect to $t$ applying Leibniz integral rule the integro-differential equation reads
\begin{eqnarray}\label{eq:integro_kernel}
 \dot{\Phi}(t) + K(0)\Phi(t) + \int_0^t \dot{K}(t-t^\prime)\Phi(t^\prime)\mathrm{d} t^\prime = \dot{f}(t).
\end{eqnarray}
Introducing $0<\bar{t}<t$ the integral can be split to separate short from long times
\begin{eqnarray}
 \fl\dot{\Phi}(t) + K(0)\Phi(t) + \int_0^{\bar{t}} \dot{K}(t-t^\prime)\Phi(t^\prime)\mathrm{d} t^\prime+ \int_{\bar{t}}^t \dot{K}(t-t^\prime)\Phi(t^\prime)\mathrm{d} t^\prime = \dot{f}(t).
\end{eqnarray}
Using integration by parts for the second integral and substituting $t^\prime$ by $t-t^\prime$ the time derivative can be moved from $K$ to $\Phi$ and we arrive at
\begin{eqnarray}\label{eq:integro_correlator}
 \fl\dot{\Phi}(t) + K(t-\bar{t})\Phi(\bar{t}) + \int_0^{\bar{t}}\dot{K}(t-t^\prime)\Phi(t^\prime)\mathrm{d} t^\prime + \int_0^{t-\bar{t}}\dot{\Phi}(t-t^\prime)K(t^\prime)\mathrm{d} t^\prime = \dot{f}(t).
\end{eqnarray}
Using the decimated time grid the first integral can be rearranged
\begin{eqnarray}
 \int_0^{\bar{t}}\dot{K}(t-t^\prime)\Phi(t^\prime)\mathrm{d} t^\prime \approx \sum_{j=1}^{\bar{i}}\left[K(i-j+1)-K(i-j)\right]\mathrm{d} \Phi(j),
\end{eqnarray}
with moments
\begin{eqnarray}
 \mathrm{d} \Phi(j)=\frac{1}{2^d h}\int_{t_{j-1}}^{t_j}\Phi(t^\prime)\mathrm{d} t^\prime,
\end{eqnarray}
and the approximation for the derivative
\begin{eqnarray}
 \dot{K}(t) \approx \frac{K(j)-K(j-1)}{2^d h}.
\end{eqnarray}
The same considerations apply for the second integral. In contrast, the time derivative $\dot{\Phi}(t)$ outside the integral is replaced by a second-order-accuracy backward finite difference formula
\begin{eqnarray}
 \dot{\Phi}(t) \approx \frac{1}{2^d h}\left[\frac{1}{2}\Phi(i-2)-2\Phi(i-1)+\frac{3}{2}\Phi(i)\right].
\end{eqnarray}
Altogether, the solution for $\Phi(i)$ is given by~\cite{Fuchs:JoP:1991}
\begin{eqnarray}
 \Phi(i)=\frac{1}{A}\left[B(i)-C(i)\right],
\end{eqnarray}
with
\begin{eqnarray}
 A = \frac{3}{2}\frac{1}{2^d h}+ \mathrm{d} K(1),\\
 B(i) = \dot{f}(i)-\frac{1}{2^d h}\left[\frac{1}{2}\Phi(i-2)-2\Phi(i-1)\right]-\mathrm{d} \Phi(1)K(i),\\
 C(i) = K(i-\bar{i})\Phi(\bar{i})+I_1^\prime+I_2^\prime,
\end{eqnarray}
and
\begin{eqnarray}
I_1^\prime = -K(i-1)\mathrm{d} \Phi(1) + \sum_{j=2}^{\bar{i}}\left[K(i-j+1)-K(i-j)\right]\mathrm{d} \Phi(j),\\
I_2^\prime = -\Phi(i-1)\mathrm{d} K(1) +\sum_{j=2}^{i-\bar{i}}\left[\Phi(i-j+1)-\Phi(i-j)\right]\mathrm{d} K(j),
\end{eqnarray}
where we take $\bar{i}=\lfloor i/2 \rfloor$ with the floor function $\lfloor x \rfloor$.

The equation of motion for the correlator, equation~\eref{eq:eom_correlator_effective}, is of the same type as equation~\eref{eq:integro_kernel} identifying $S_\mu(q,t)$ with $\varphi(t)$ and using the commutativity of the convolution. By splitting the integral and using integration by parts it can be rearranged
\begin{eqnarray}
 \fl\dot{\Phi}(t) +\Gamma \Phi(t)+K(t-\bar{t})\Phi(\bar{t})-K(t)\Phi(0) &+\int_0^{\bar{t}}\dot{K}(t-t^\prime)\Phi(t^\prime)\mathrm{d} t^\prime \nonumber\\ &+ \int_0^{t-\bar{t}}\dot{\Phi}(t-t^\prime)K(t^\prime)\mathrm{d} t^\prime = 0,
\end{eqnarray}
with the same solution strategy as described above. The right hand side vanishes, which is equivalent to $\dot{f}(i)=0$. Due to the additional summand $\Gamma \Phi(t)$ in the preceding equation, $\Gamma$ is added to the parameter $A$. The quantity $-K(t)\Phi(0)$ reflects in a further term $\Phi(0)K(i)$ within $B(i)$.

\section{ISF for longer simulation times\label{sec:appendix_ISF}}
\begin{figure}[t]
 \centering
  \includegraphics{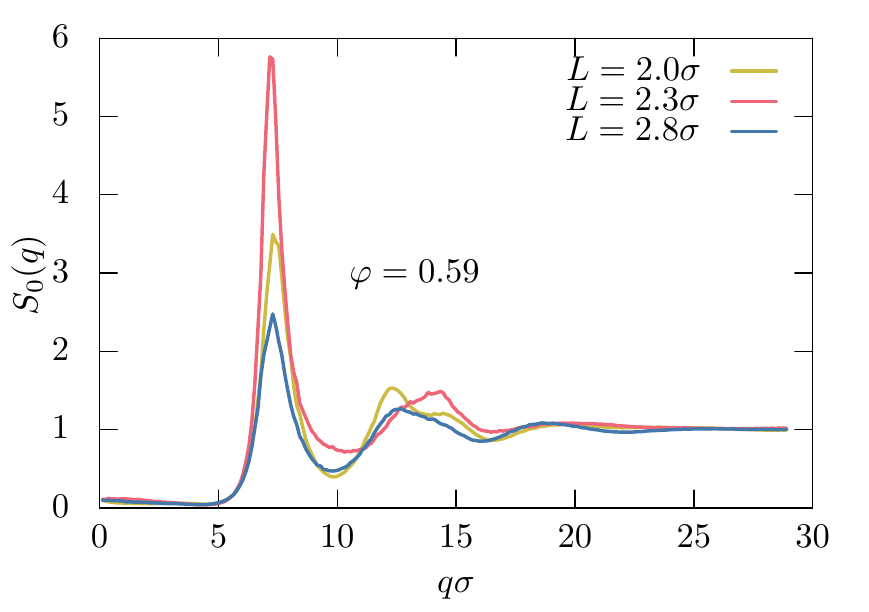}
  \caption{Static structure factors $S_0(q)$ for different values of $L$ at packing fraction $\varphi=0.59$ for longer sample preparation.}
  \label{fig:structure_eq_time}
\end{figure}

To ensure that the conclusions drawn are not affected by aging of the system, 
we have repeated the simulations presented in the main text with a longer sample-preparation  time of $2\times10^4t_0$ (20 times longer than used originally) before taking measurements. 
 The results of these simulations have been analyzed in the same way as described in the main text. We see that after longer sample preparation the static structure factor has a  slightly higher first peak for all confinement lengths, \Fref{fig:structure_eq_time}, but there is no change in the shape of the curves.

\begin{figure}[t]
 \centering
  \includegraphics{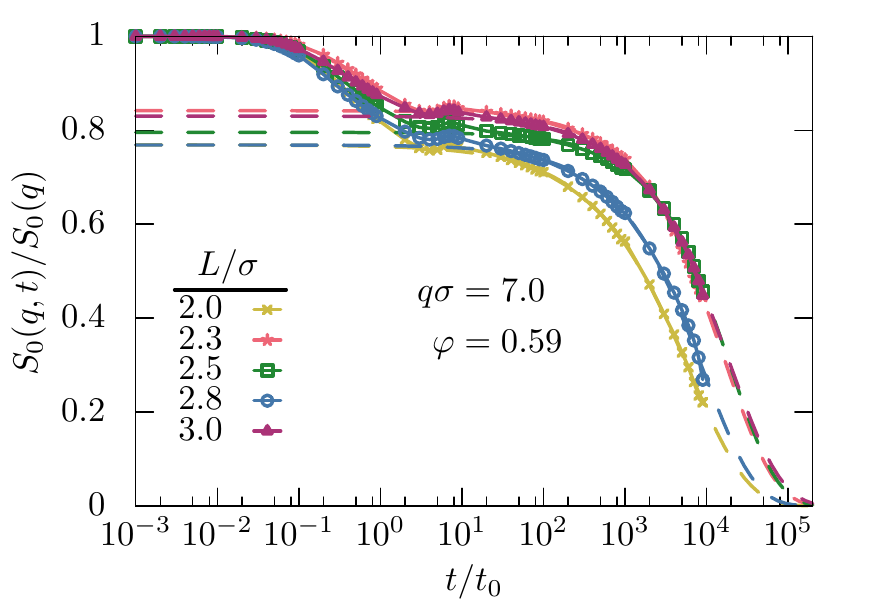}
\caption{Normalized ISF $S_0(q,t)/S_0(q)$ for $q\sigma=7.0$ at packing fraction $\varphi=0.59$ for longer simulations.  The dashed line are  KWW fits.}
\label{fig:ISF_eq_time}
\end{figure}

The normalized ISF, \Fref{fig:ISF_eq_time}, shows the same trend with increasing confinement length $L$ as for the shorter  sample preparation, \Fref{fig:ISF}(b). 
We observe the nonmonotonic behavior clearly between the strongest confinement length, $L=2.0\sigma$, which shows the fastest decay of correlations, to an intermediate value, $L=2.3\sigma$, which has the slowest relaxation, followed by faster relaxation again at a further increased confinement length, $L=2.8$, exactly as we saw for the  shorter sample preparation. The main difference in increasing the sample-preparation time is 
that the relaxation time for all confinement lengths increase.

\begin{figure}[t]
 \centering
  \includegraphics{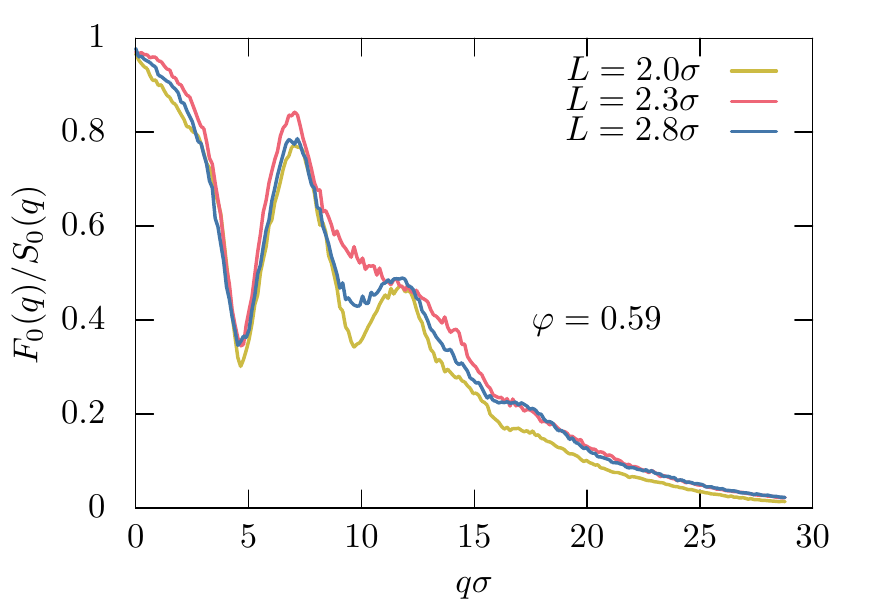}
\caption{Normalized  nonergodicity parameters $F_0(q)/S_0(q)$ as a function of the wavenumber $q$ for different values of $L$ at packing fraction $\varphi=0.59$ for longer sample preparation.}
\label{fig:nonergodicity_eq_time}
\end{figure}

The wavelength dependence of the nonergodicity parameters, \Fref{fig:nonergodicity_eq_time} also look practically indistinguishable from the shorter initial  simulation time, \Fref{fig:nonergodicity}(b), and the
 same nonmonotonic trends with confinement length are observed.

Figures \ref{fig:structure_eq_time} -  \ref{fig:nonergodicity_eq_time} demonstrate that the conclusions drawn in the main body of this manuscript are not dependent on the sample-preparation time used in 
the simulations. Despite this, the longer relaxation times observed in the ISF, \Fref{fig:nonergodicity_eq_time}, for the simulations with a longer sample preparation indicate that our system is aging. It is expected from previous studies in bulk that polydisperse mixtures of hard spheres
 at high packing fractions demix and then crystallize \cite{Lindquist:JCP:2018, Bommineni:PRL:2019}. This effect is enhanced in hard-sphere mixtures confined between walls, where the 
crystallization and demixing occur at packing fractions below the glass-transition line~\cite{Jung:2020_2}. In fact, if we simulate our quasi-confined system with accelerated dynamics, 
using the SWAP Monte-Carlo algorithm \cite{Berthier:PRL:2016,Ninarello:PRX:2017}, we find evidence of demixing, as indicated by a strong low-$q$ peak in the partial structure factor 
(the structure factor of only the smallest particles, those with diameter $<0.9\sigma$ )~\cite{Ninarello:PRX:2017}. This is also evident from the full structure factor, where a significant peak at very 
low $q$ can be observed. This indicates that even with the normal molecular-dynamics algorithm used in the body of this manuscript, our system would eventually demix. However, the use of the SWAP 
algorithm essentially allows access to time scales much longer than would be accessible in a relevant experiment, for example, a colloidal glass~\cite{Ninarello:PRX:2017}, and so this 
demixing is likely to be irrelevant to the glass transition of approximately hard-sphere colloids. Since the goal of this work is to provide deeper insight into the effect of confinement on the glass 
transition in such systems, we restrict our analysis to relatively short time scales, where demixing and crystallization does not occur yet, and the dynamics are glassy. In both simulations, the ones for  short sample preparation
 presented in the main text, and the others for longer sample preparation considered now, we see no low $q$ peak in the partial structure factor, indicating that the particles have not demixed.

Even at lower packing fraction ($\varphi=0.57$), where the dynamics are not glassy and there is no real plateau in the ISF, the particles demix when simulated with accelerated SWAP dynamics. Additionally, we find that this phenomenon is not sensitive to the specifics of the polydispersity. We observe demixing also for mixtures of hard spheres which have a Gaussian or tophat particle-size distribution. As such, we conclude that the best option for studying glassy dynamics of quasi-confined hard spheres is to start with a homogenous mixture of polydisperse particles and use sufficiently short simulations that demixing and crystallization do not play a role, as we have done here.

\section*{References}
\bibliography{dynamics_quasi_confinement}

\end{document}